\documentclass[9pt,twocolumn,twoside]{optica-mod}
\setboolean{shortarticle}{false}
\setboolean{minireview}{false}
\title{Stably accessing octave-spanning microresonator frequency combs in the soliton regime}

\author[1,3,*]{Qing Li}
\author[2]{Travis C. Briles}
\author[1]{Daron A. Westly}
\author[2,4]{Tara E. Drake}
\author[2,4]{Jordan R. Stone}
\author[1]{B. Robert Ilic}
\author[2,4]{Scott A. Diddams}
\author[2]{Scott B. Papp}
\author[1,*]{Kartik Srinivasan}

\affil[1]{Center for Nanoscale Science and Technology, National Institute of Standards and Technology, Gaithersburg, Maryland 20899, USA}
\affil[2]{Time and Frequency Division, National Institute of Standards and Technology, Boulder, Colorado 80305, USA}
\affil[3]{Maryland NanoCenter, University of Maryland, College Park, Maryland 20742, USA}
\affil[4]{Department of Physics, University of Colorado, Boulder, Colorado 80309, USA}

\affil[*]{Corresponding author: qing.li@nist.gov, kartik.srinivasan@nist.gov}

\dates{Compiled \today}

\begin{abstract}
Microresonator frequency combs can be an enabling technology for optical frequency synthesis and timekeeping in low size, weight, and power architectures. Such systems require comb operation in low-noise, phase-coherent states such as solitons, with broad spectral bandwidths (e.g., octave-spanning) for self-referencing to detect the carrier-envelope offset frequency. However, stably accessing such states is complicated by thermo-optic dispersion.  For example, in the Si$_3$N$_4$ platform, precisely dispersion-engineered structures can support broadband operation, but microsecond thermal time constants have necessitated fast pump power or frequency control to stabilize the solitons. In contrast, here we consider how broadband soliton states can be accessed with simple pump laser frequency tuning, at a rate much slower than the thermal dynamics.  We demonstrate octave-spanning soliton frequency combs in Si$_3$N$_4$ microresonators, including the generation of a multi-soliton state with a pump power near 40 mW and a single-soliton state with a pump power near 120 mW. We also develop a simplified two-step analysis to explain how these states are accessed in a thermally stable way without fast control of the pump laser, and outline the required thermal properties for such operation. Our model agrees with experimental results as well as numerical simulations based on a Lugiato-Lefever equation that incorporates thermo-optic dispersion. Moreover, it also explains an experimental observation that a member of an adjacent mode family on the red-detuned side of the pump mode can mitigate the thermal requirements for accessing soliton states.
\end{abstract}

\setboolean{displaycopyright}{false}
\begin{document}

\maketitle

\section{Introduction}
Soliton states in Kerr microcavities represent a path to low-noise comb formation with properties suitable for metrological applications \cite{Kippenberg_comb_nature2007, Kippenberg_comb_science2011}, including optical frequency synthesis \cite{Russell:2000, Metrology_comb:2002}, optical clocks \cite{Diddams825, Papp_clock_Optica}, microwave generation \cite{Diddams_microwave_link, Vahala_Soliton}, etc. However, accessing soliton states in an experiment is challenging due to the complicating presence of thermo-optic dispersion that makes stable operation on the red-detuned side of the cavity resonance difficult \cite{Kippenberg_soliton_2014}. So far soliton generation has been demonstrated in MgF$_2$ \cite{Kippenberg_soliton_2014}, SiO$_2$ \cite{Diddams_phase_step, Vahala_Soliton}, Si \cite{Gaeta_soliton_Si}, and Si$_3$N$_4$~\cite{Gaeta_soliton_Si3N4,Wong_soliton_Si3N4, Weiner_soliton_Si3N4, Kippenberg_DW_science} microresonators, but the reported approaches for stably accessing these states in the literature are quite different, including various frequency ramping schemes such as forward and backward pump laser scans \cite{Kippenberg_soliton_2014, Gaeta_thermal_tuning_soliton_Si3N4, Weiner_thermal_tuning_comb, Kippenberg_backward_tuning}, pump power modulation \cite{Papp_seeding_OE_2013, Diddams_injection_locking}, as well as abrupt power changes over timescales that are faster than the thermal lifetime (power kicking)~\cite{Kippenberg_power_kicking, Vahala_power_kicking}. While there are some rough guidelines available for each demonstrated method, they have not offered a clear, systematic path to overcoming the underlying thermal challenges and adiabatically reaching the soliton regime. Moreover, none of these works have reached the regime of octave-spanning operation, a pre-requisite for the $f$-$2f$ self-referencing technique for determining the carrier-envelope offset frequency \cite{Diddams_microwave_link, Telle1999, Diddams_self_ref}.

Here, we present the first experimental demonstration (to the best of our knowledge) of phase-coherent, octave-spanning soliton microcomb states in Si$_3$N$_4$ microresonators. We access these soliton states through the slow frequency tuning of a fixed power pump laser, at a rate that is much slower than the thermal (and Kerr) dynamics. To better understand our system, we present a simplified analysis to systematically study the thermal stability of the available soliton states, and the results agree with the experimental data and full numerical simulations that consider both the Kerr and thermal effects. This method also allows us to explain an interesting experimental observation that an adjacent mode family member on the red-detuned side of the pump resonance helps stabilize the thermal dynamics during soliton formation, which represents a novel means to generate solitons with slow pump frequency tuning and without a significant improvement to the resonator quality factor, absorption rate, or thermal conductance.

\section{Overview}
\label{sec:overview}

\begin{figure}[t]
	\centering
	\includegraphics[width=\linewidth]{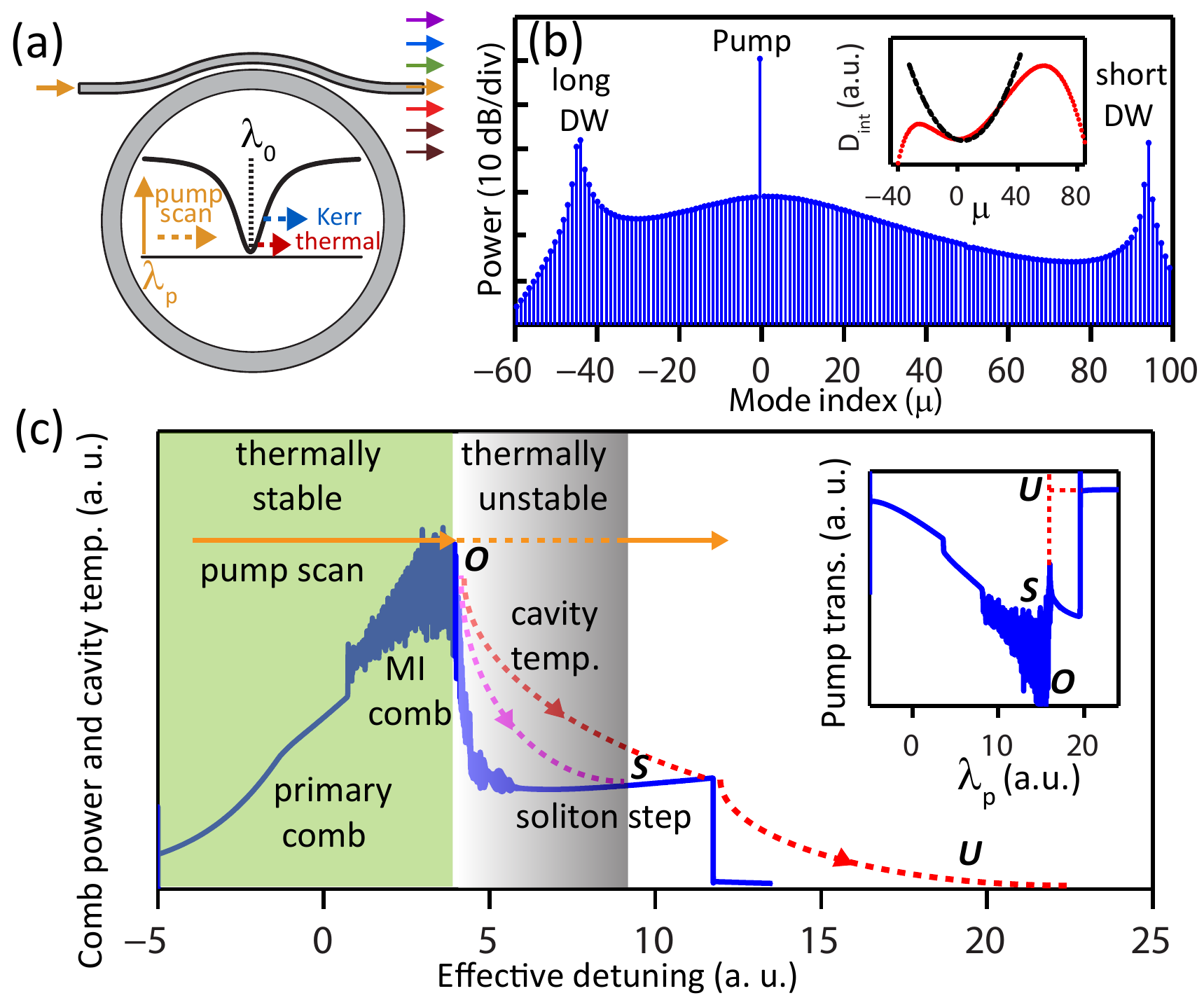}
	\caption{Overview. (a) Schematic of a waveguide-coupled microring resonator in which a pump laser at wavelength $\lambda_{p}$ is slowly tuned across a cavity mode whose cold-cavity wavelength is $\lambda_{0}$. (b) Simulated steady-state soliton microcomb spectrum for a dispersion profile shown in the inset, as a function of mode index $\mu$, where $\mu$=0 is the pumped mode (at 1550~nm) and the mode spacing is $\approx$~1 THz. Near $\mu$=0, the dispersion is quadratic (see the inset). The departure from quadratic dispersion away from $\mu$=0 results in dispersive waves (DWs) near 1 $\mu$m and 2 $\mu$m. (c) Comb power as a function of effective detuning (blue line) as the pump laser is tuned in a wavelength-increasing direction at a slow enough rate (detailed definitions are given in Section \ref{sec:thermal_stability_model}). At the crossover between MI combs and soliton combs (point $\textbf{O}$, discussed in more detail in Section~\ref{sec:thermal_stability_model}), the cavity temperature becomes dynamically unstable.  In certain cases, it stabilizes with the system in a soliton state (dashed magenta line with the end state at point $\textbf{S}$), while in other cases it drops to that of the ambient environment and no soliton state is accessed (dashed red line with the end state at point $\textbf{U}$). The inset shows the corresponding pump transmission for these two cases. Note that the transition from $\mathbf{O}$ to $\mathbf{S}$ or $\mathbf{U}$ happens on the order of the thermal lifetime and is difficult to resolve if the data acquisition rate is not fast enough.}
\label{fig:overview}
\end{figure}

Figure~\ref{fig:overview} summarizes our understanding of how to access octave-spanning microcomb soliton states in a thermally stable way through slow frequency tuning of the pump laser. Soliton states exist on the red-detuned side of the Kerr-shifted pump resonance~\cite{Kippenberg_soliton_2014}, but fixing the pump laser wavelength $\lambda_{p}$ at such a detuning is complicated by thermal effects, which cause frequency shifts that depend on $\lambda_p$ and the intracavity power and can vary over relatively fast timescales.  While for a resonator with an appropriate group velocity dispersion profile, octave-spanning soliton states such as that shown in Fig.~\ref{fig:overview}(b) are a solution of the governing Lugiato-Lefever equation (LLE)~\cite{LLE_chembo,LLE_Cohen}, the path required to access such states experimentally requires the LLE to be supplemented by thermal considerations.

Our model involves a careful thermal stability analysis for different comb states obtained from the LLE method. For example, the blue trace in Fig.~\ref{fig:overview}(c) shows one example of numerical simulations based on the pure LLE model (no thermal effects), illustrating the three main stages of comb generation: primary comb, modulation-instability (MI) comb, and the soliton regime~\cite{Kippenberg_soliton_2014}. The primary comb and MI comb states are thermally stable, as they are on the blue-detuned side of the Kerr-shifted pump resonance, where the thermo-optic effect provides a negative feedback to the cavity energy. In contrast, soliton states exist on the effectively red-detuned side of the Kerr-shifted resonance, and are typically accompanied by a large drop in average intracavity power relative to the MI states.  Thus, in the transition from an MI comb to a soliton state, the cavity temperature become dynamically unstable (point $\textbf{O}$) and drops at a rate determined by the thermal lifetime (on the order of microseconds for the Si$_3$N$_4$ resonators considered in this work), which in turn brings the cavity into a further red-detuned regime. As a result of the vastly different timescales of the thermal response and the laser sweep, the laser frequency can be considered fixed as the temperature changes, so that the effective cavity detuning is largely dictated by the resonator thermal response.

\begin{figure*}[t]
	\centering
	\includegraphics[width=0.9\linewidth]{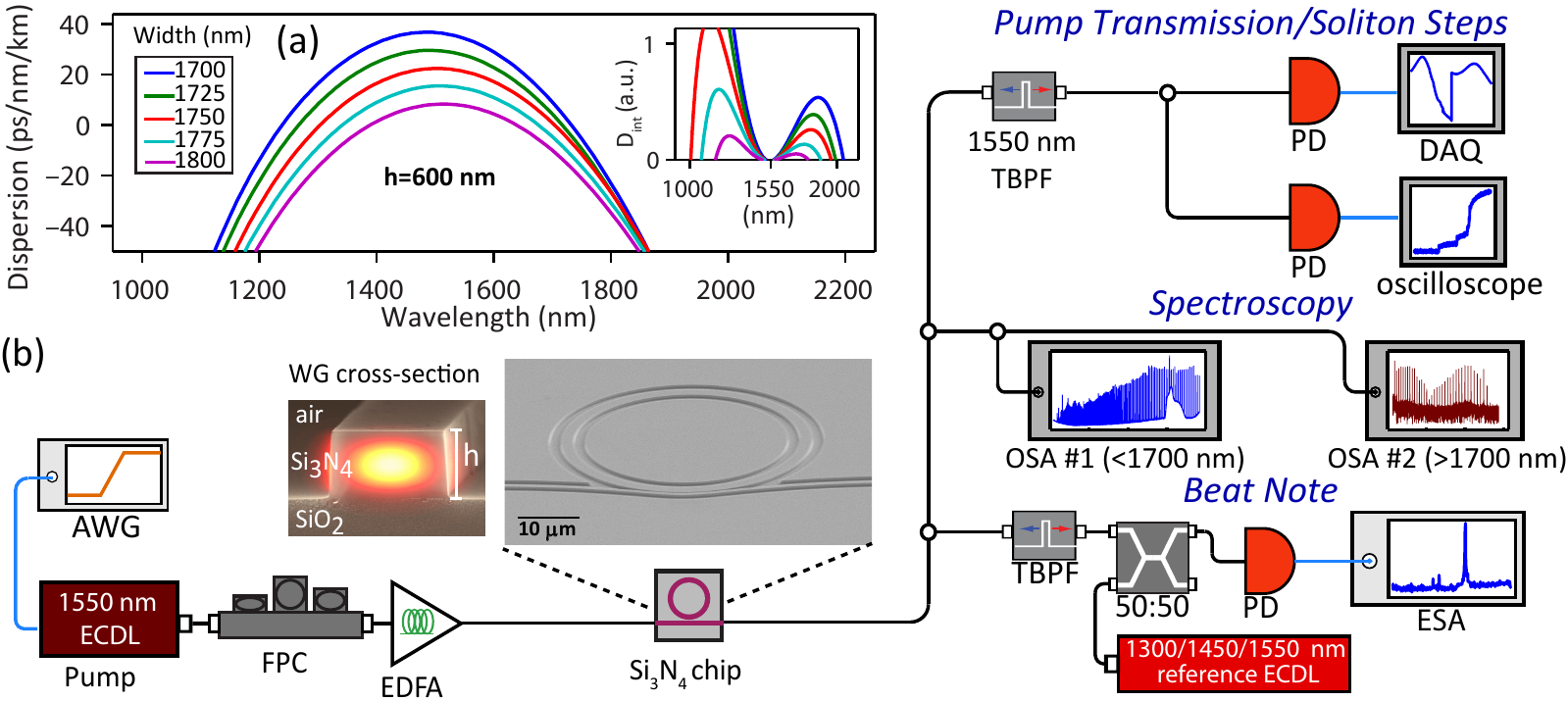}
	\caption{Si$_3$N$_4$ microring dispersion and experimental setup. (a) Calculated dispersion of a 23 $\mu$m radius Si$_3$N$_4$ microring for a fixed thickness of 600 nm and varied ring widths. The inset shows the corresponding integrated dispersion. (b) Experimental setup. AWG, arbitrary waveform generator; ECDL, external cavity diode laser; FPC, fiber polarization rotator; EDFA, erbium-doped fiber amplifier; TBPF, tunable bandpass filter; PD, photodetector; DAQ, data acquisition; OSA, optical spectrum analyser; ESA, electronic spectrum analyzer. The zoom-in figure of the chip shows a scanning-electron micrograph of the microring.}
	\label{fig:setup}
\end{figure*}

The red and magenta curves in Fig.~\ref{fig:overview}(c) depict two examples of the evolution of the laser-cavity relative detuning after passing into the soliton regime. For a thermally stable soliton state (dashed magenta line in Fig.~\ref{fig:overview}(c)), the cavity temperature stabilizes with the system at point $\textbf{S}$, and the system will remain in this state as we continue varying the pump wavelength, until the pump laser is completely off resonance. In the thermally unstable case (dashed red line in Fig.~\ref{fig:overview}(c)), the thermal trajectory of the cavity fails to stabilize with the system on any soliton state, and the cavity temperature drops to the ambient temperature, so that the pump laser is automatically detuned off resonance (point $\textbf{U}$). These two cases can be easily distinguished in the pump transmission measurement if we vary the pump wavelength at a slow enough rate relative to the thermal dynamics. As illustrated by the inset of Fig.~\ref{fig:overview}(c), a thermally stable soliton results in a characteristic soliton step in the pump transmission, which is absent if no such thermally stable soliton states exist.  In the following sections, we present experimental observations and the results of numerical simulations to elucidate these scenarios.

The organization of this manuscript is as follows.  Sections~\ref{sec:dispersion_and_setup} and ~\ref{sec:multi_soliton_expt} present the dispersion design, measurement setup, and experimental observation of octave-spanning multi-soliton states. Section~\ref{sec:thermal_stability_model} gives a detailed development of the model summarized above, and describes how the thermally stable solutions can be predicted based on a simple graphical analysis incorporating solutions of the individual LLE and thermo-optic models.  Section~\ref{sec:thermal_stability_sims} compares the results of this model against simulations in which thermal dynamics are fully incorporated into the LLE. Finally, Section~\ref{sec:single_solitons} presents the observation of octave-spanning single solitons, and analyzes how this observation is enabled by a mode coupling effect that thermally stabilizes the system.

\section{Dispersion design and experimental setup}
\label{sec:dispersion_and_setup}

Our focus is on Si$_3$N$_4$ microring resonators whose geometries are tailored for octave-spanning comb generation. The key element is a 23 $\mu$m radius Si$_3$N$_4$ microring with 1 THz free spectral range, as the large mode spacing allows comb generation with reduced pump powers compared to larger-sized microresonators \cite{Gaeta_octave,Gaeta_soliton_Si3N4,Kippenberg_DW_science}.  The generated comb spectrum depends critically on the ring waveguide dispersion~\cite{Gaeta_octave,Wang_broadband_design}, and in particular, broad spectral bandwidths can be achieved by soliton dispersive wave emission~\cite{Kippenberg_DW_science}.  To determine our resonator dispersion, we must consider the bending effect from the small ring outer radius (23~$\mu$m), which shifts the resonator towards the normal dispersion regime, requiring smaller ring widths to achieve a given level of anomalous dispersion than what is needed for larger radius resonators (where bending effects can be neglected). The overall dispersion, which includes the material dispersion as determined by spectroscopic ellipsometry, is computed based on a fully vectorial microresonator eigenfrequency mode solver, and a series of examples is provided in Fig.~\ref{fig:setup}(a). For air-clad Si$_3$N$_4$ microrings (see Fig.~\ref{fig:setup}(b) for waveguide cross-section), anomalous dispersion can be attained for Si$_3$N$_4$ thickness around 600 nm, which is considerably less than what is typically required for oxide-clad devices. For the ring widths shown in Fig.~\ref{fig:setup}(a), the second-order dispersion is anomalous for a wavelength range spanning a few hundred nanometers and displays a monotonic decrease with the ring width. Moreover, high-order dispersion terms are found to be important when designing frequency combs for such wide spectral ranges \cite{Wang_broadband_design}. In particular, they are responsible for the aforementioned generation of dispersive waves, which are coherently linked to the pump and can extend the comb spectrum to the normal dispersion regime \cite{Kippenberg_DW_science}. The spectral positions of these dispersive waves correspond to the zero-crossing points of the integrated dispersion relative to the pumped mode, which is given by
\begin{align}
D_\text{int}(\mu)&\equiv \omega_\mu -(\omega_0+D_1\mu) \notag \\
&= \frac{1}{2!}D_2\mu^2+ \frac{1}{3!}D_3\mu^3+\frac{1}{4!}D_4\mu^4\dots,  \label{eqn:Dint}
\end{align}
where $\mu$ is an integer representing the relative mode number, $\omega_\mu$ is the resonance frequency of the $\mu_\text{th}$ mode ($\mu=0$ is the pump), $D_1$ is the free spectral range of the microresonator, $D_2$ is the second order dispersion, and $D_3,D_4\dots$ are higher-order dispersion terms. For Si$_3$N$_4$ thickness around 600 nm, numerical simulations (see the inset of Fig.~\ref{fig:setup}(a)) indicate that for ring widths in the range of 1750 nm to 1800 nm, there are two dispersive waves located around wavelengths of 1 $\mu$m and 2 $\mu$m \cite{Kartik_comb_FiO}, while for ring widths with stronger dispersions (width < 1750 nm) only the long dispersive wave around 2 $\mu$m is observed within this spectral window. Finally, we have used the pulley-coupling scheme to achieve an efficient power injection for the pump as well as an optimized out-coupling for the frequency components around the wavelengths of 1 $\mu$m and 2 $\mu$m \cite{Kartik_FWMBS, Kartik_comb_FiO}.

\begin{figure*}[t]
	\centering
	\includegraphics[width=0.86\linewidth]{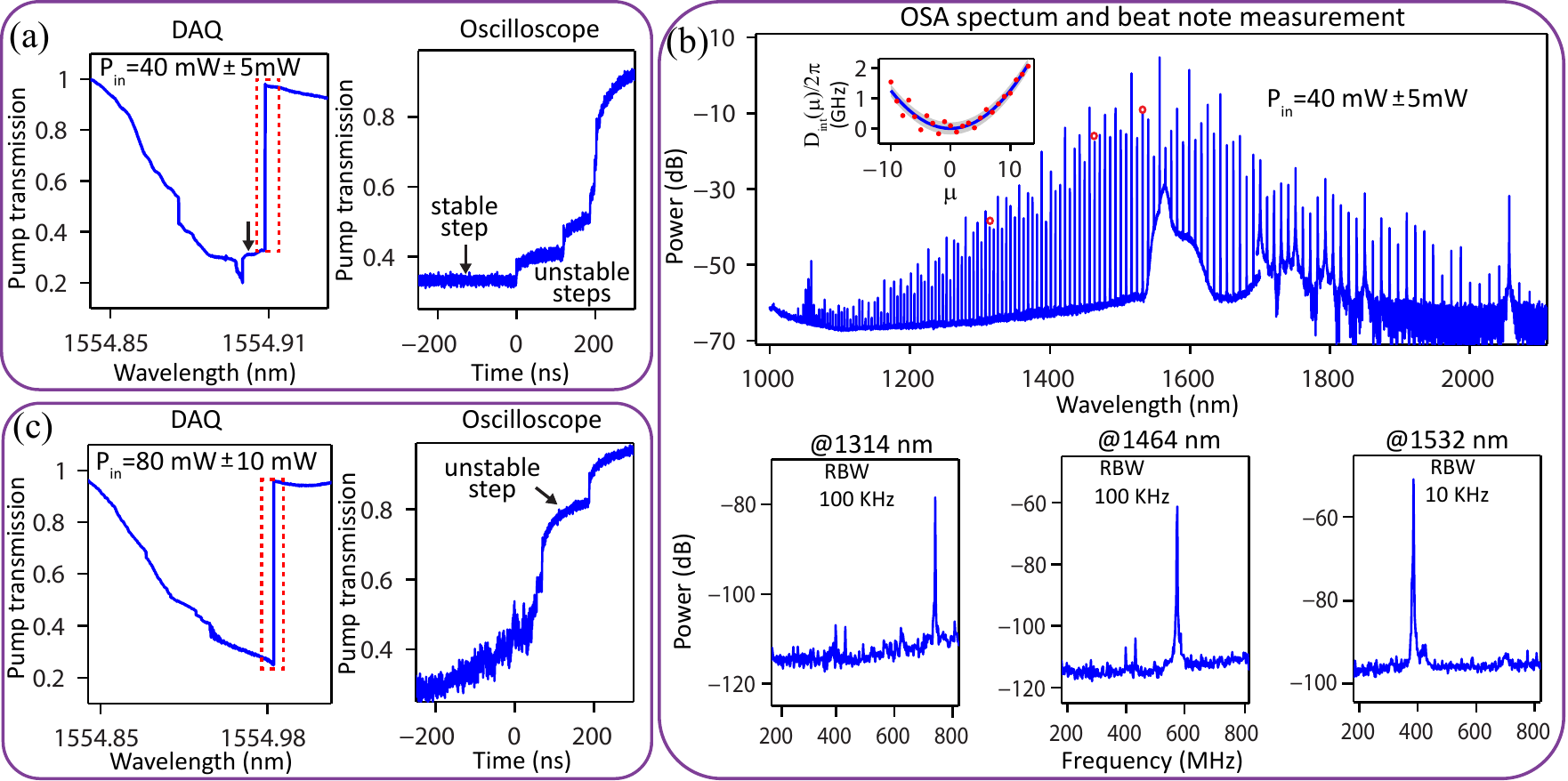}
	\caption{Experimental results from a Si$_3$N$_4$ microring with width of 1760 nm $\pm$ 10 nm and height of 603 nm $\pm$ 2 nm. (a) Pump transmission for an input power of 40 mW $\pm$ 5 mW at a pump tuning speed of $-100$ GHz/s. The oscilloscope data corresponds to the boxed region in the DAQ transmission, and thermally stable and unstable soliton steps are labeled. (b) The upper figure shows the comb spectrum corresponding to the stable soliton step with the pump detuning marked by the vertical arrow in (a). This spectrum includes output fiber coupling loss ($\approx$6~dB for the pump wavelength). The inset shows the measured resonance dispersion (red dots) versus a quadratic fit (blue solid line), and D$_2/2\pi$ is extracted to be 25 MHz $\pm$ 2 MHz. The lower figures are beat-note measurements for the three selected comb lines from the above comb spectrum (marked by red circles in the spectrum). The linewidths of the beat-note signals are a few MHz and are mostly limited by the frequency stability of the pump laser and the reference laser. Note that the additional small spikes in the ESA spectrum are confirmed to be electronic noise and persist even when the optical power is turned off. A power of 0 dB is referenced to 1 mW (i.e., dBm). RBW: resolution bandwidth of the ESA. (c) Pump transmission for the input power of 80 mW $\pm$ 10 mW. The oscilloscope data corresponds to the boxed region in the DAQ transmission.}
	\label{fig:exp_H600nm}
\end{figure*}

Figure~\ref{fig:setup}(b) shows the experimental setup. Light is coupled on and off the chip with lensed fibers, with a coupling loss of approximately 6 dB per facet. At the output, we obtain the pump transmission using a narrowband filter that is centered on the pump mode. In this work, the pump frequency laser is scanned at a relatively slow speed, with a full resonance scan that typically takes up to tens of milliseconds, corresponding to a pump tuning rate on the order of -100 GHz/s. During the sweep, however, the cavity mode becomes dynamically unstable at certain pump detunings due to the thermo-optic effect~\cite{Vahala_thermal}, which quickly shifts the resonant mode off the pump laser at a timescale proportional to the thermal lifetime ($\approx 3.4~\mu$s, see Supplementary Material). To measure these distinctively different temporal responses, a photoreceiver with 800 MHz bandwidth is used in combination with a data acquisition card (DAQ) for the slow response (sub-millisecond), and with an oscilloscope (600 MHz bandwidth) for the fast response (sub-microsecond). In addition, because of the wide spectral span of the generated frequency comb, the transmitted light is sent to two optical spectrum analyzers (OSAs), with one covering the wavelength range of 600 nm to 1700 nm and the other covering 1700 nm to 2600 nm. Finally, the coherence properties of the generated frequency combs are characterized by measuring the beat note between individual comb lines and reference lasers tuned into nearly the same frequencies. After combining them using a 50-50 directional coupler, their beat note is obtained with the use of the same 800 MHz bandwidth photoreceiver for the pump transmission and an electronic spectrum analyzer. In these measurements, a coherent comb state corresponds to a single narrow beat note whose linewidth is only limited by the laser linewidth, while a noisy comb state typically results in either multiple beat notes or a broad beat note \cite{Comb_coherence_Cohen}.

\section{Accessing multi-soliton states: experiment}
\label{sec:multi_soliton_expt}

We next present experimental results on octave-spanning, phase coherent microcomb states, accessed through tuning of the pump laser frequency at a rate much slower than any time constant of the system, and consistent with multi-soliton behavior.

Figure~\ref{fig:exp_H600nm} shows the experimental results for a Si$_3$N$_4$ microring with thickness of 603 nm $\pm$ 2 nm and width of 1760 nm $\pm$ 10 nm, where the uncertainty in thickness is a one standard deviation value based on ellipsometry of the grown film, and the uncertainty in width is a one standard deviation value based on scanning electron microscopy. We start with a relatively small pump power (40 mW $\pm$ 5 mW on chip) and slowly decrease the pump frequency\footnote{The uncertainty in pump power is a one standard deviation value and is from the variation in chip insertion loss across multiple devices}. As shown by the DAQ response in Fig.~\ref{fig:exp_H600nm}(a), the pump transmission initially varies with frequency in a standard way, and a broadened lineshape is recorded due to the combined Kerr and thermal effects. In addition, a step response indicative of soliton generation is observed near the bottom of the pump transmission \cite{Kippenberg_soliton_2014}. Its spectral width is independent of the laser scanning speed (provided that it is slow enough), suggesting that it is a thermally stable state. As we further reduce the pump frequency, however, thermally-induced instability spontaneously shifts the resonant mode off the pump laser in a time on the order of microseconds (boxed region in Fig.~\ref{fig:exp_H600nm}(a)). The oscilloscope data (Fig.~\ref{fig:exp_H600nm}(a)) reveals additional soliton steps during this fast transition.  However, these soliton steps are thermally unstable and only represent transient states.

The optical spectrum corresponding to the thermally stable soliton step is provided in Fig.~\ref{fig:exp_H600nm}(b), and shows an octave-spanning frequency comb with two dispersive waves. Their spectral positions, with the short and long dispersive waves located around $1.06\ \mu$m and $2.06\ \mu$m, respectively, are consistent with predictions based on a similar dispersion calculation as shown in Fig.~\ref{fig:setup}(b) for the estimated ring waveguide dimension (note that the spectral positions of the dispersive waves shift toward longer wavelengths as the Si$_3$N$_4$ thickness increases). The second-order dispersion is also estimated through measurement of the resonance frequencies using a wavemeter in the $1.55\ \mu$m band, which gives $D_2/2\pi=25$ MHz $\pm$ 2 MHz (see the inset of the comb spectrum in Fig.~\ref{fig:exp_H600nm}(b)), where the uncertainty is due to the resolution of the wavemeter and is a one standard deviation value from the fit to the data. Finally, the coherence properties of the generated comb are examined with the help of reference lasers, each of which has a linewidth $< 1$ MHz. The obtained clean beat-note signals for three selected comb lines (see the bottom of Fig.~\ref{fig:exp_H600nm}(b)) confirm that the comb spectrum shown in Fig.~\ref{fig:exp_H600nm}(b) corresponds to a phase-coherent state.

\begin{figure*}[t]
	\centering
	\includegraphics[width=0.9\linewidth]{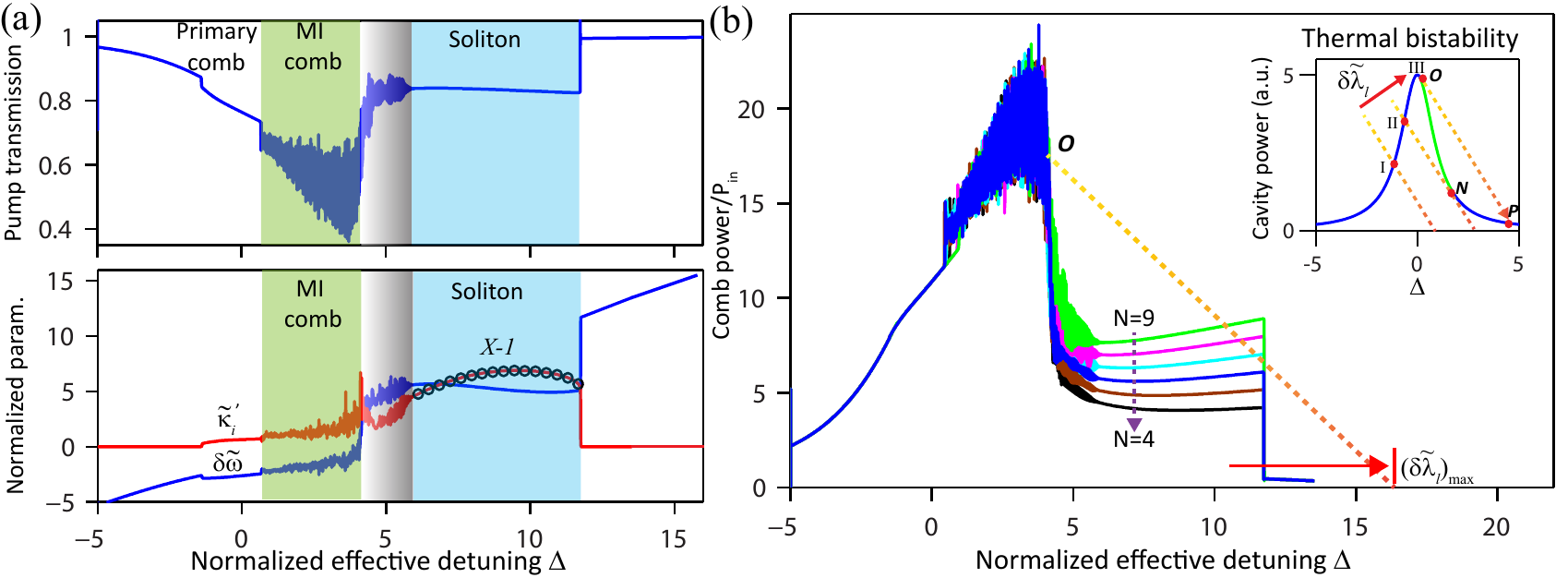}
	\caption{Illustration of the comb generation dynamics and two-step analysis to study the thermal stability of soliton states. (a) Simulation results for the comb generation dynamics in a Si$_3$N$_4$ microresonator, based on the LLE model without thermal effects. The upper figure is for the pump transmission and the lower figure is for the normalized parameters $\tilde{\kappa_i'}$ (red line) and $\delta\tilde{\omega}$ (blue line), which are the effective loss rate due to conversion from pump to comb lines and the detuning with respect to the Kerr-shifted pump resonance, respectively. The markers (black circles) denote the calculated value of $X-1$, with $X$ being the ratio of the intracavity power of the whole comb spectrum to that of the pump comb line. (b) Simulated comb powers for repeated runs of the LLE simulation using the same parameters in (a). $\mathbf{O}$ is the point at which the slope of the tangent to the averaged comb power is equal to $-K_\text{eff}$. For a given $K_{\text{eff}}$, the maximum nonlinear wavelength shift for which a certain soliton state (here $N=6$) is thermally stable is given by $(\delta\tilde{\lambda_l})_\text{max}$. The inset depicts the case without the Kerr effect (standard thermo-optic bistability), with details provided in the text.}
	\label{fig:thermal}
\end{figure*}

Following the successful soliton generation at a pump power near 40 mW, we gradually increase the power and monitor the change in comb dynamics. As shown in Fig.~\ref{fig:exp_H600nm}(c), the soliton step near the bottom of the pump transmission disappears at a power of 80 mW $\pm$ 10 mW. In fact, the oscilloscope measures a noisier signal than the low-power case, indicating that the corresponding comb state is likely to be chaotic. On the other hand, as the resonant mode is being shifted off the pump laser by the thermo-optic effect, a low-noise soliton step is observed for a pump transmission well above its minimum. Compared to the multi-soliton state shown in Fig.~\ref{fig:exp_H600nm}(a), numerical simulations indicate that it corresponds to a lower soliton number state. However, experimentally we find it is difficult to land on this state using slow pump frequency tuning. Qualitatively, this behavior might be expected based on the large change in intracavity comb power, and resulting thermal effect, associated with soliton steps that are well above the pump transmission minimum. We consider such points in detail in the next section.

\section{Thermal stability: Modeling and discussion}
\label{sec:thermal_stability_model}

In the previous section, we showed that it is possible to stably access a multi-soliton state with two dispersive waves at a relatively small pump power (40 mW $\pm$ 5 mW). Ideally, we would also have access to a lower soliton number state, which is expected to have a broader spectral bandwidth and smoother spectral envelope.  Increasing the pump power to 80 mW $\pm$ 10 mW reveals the existence of such a state (the multi-soliton state is chaotic at this pump power), but it is thermally unstable and cannot be accessed with slow pump frequency tuning. In this section, we focus on the understanding of these experimental results with the aid of numerical simulations. To simplify this complicated problem, we shall consider a parameter regime in which the thermal effect is slow compared to the Kerr dynamics. Such an assumption, justified by our expected thermal lifetime (see Supplementary Material), allows us to decouple these two effects and greatly simplifies the analysis, as shown below.

It is well-known that the comb dynamics can be described by a mean-field Lugiato-Lefever equation (LLE) as \cite{LLE_chembo, LLE_Cohen,LLE_Gaeta_route}:
\begin{align}
\frac{\partial E(t,\tau)}{\partial t}=&\left[-\frac{\kappa}{2}-i\delta \omega_\text{eff}
        + iv_g\sum_{k\geq2} \frac{\beta_k}{k!}\left(i\frac{\partial}{\partial\tau}\right)^k + i\gamma v_g|E(t,\tau)|^2\right] \notag\\
        &\times E(t,\tau)+ i\sqrt{\kappa_c/t_{R}} E_\text{in}, \label{eq:LLE}
\end{align}
where $|E(t,\tau)|^2$ is the power travelling inside the microring ($t$ is the slow time of the cavity, associated with the resonator round-trip time $t_{R}$ and $\tau$ is the time), $\delta\omega_\text{eff}$ is the effective cavity detuning ($\delta\omega_\text{eff} \equiv \omega_p - \omega_l$ with $\omega_p$ and $\omega_l$ being the hot-cavity resonance frequency and the pump laser frequency, respectively), $v_g$ is the group velocity, $\beta_k$ is the $k$th order dispersion coefficient, $\gamma$ is the Kerr nonlinear coefficient, $\kappa_c$ is the external coupling rate ($\kappa_c=\omega_p/Q_c$ with $Q_c$ being the coupling $Q$), $\kappa$ is the total cavity loss rate ($\kappa=\kappa_i + \kappa_c$, where $\kappa_i=\omega_p/Q_i$ is the intrinsic loss rate with $Q_i$ being the intrinsic $Q$), and $E_\text{in}$ is the driving field (input power $P_\text{in}=|E_\text{in}|^2$).

We first consider the evolution of the pump transmission through the Kerr resonator. In steady-state, the pump transmission $t_p$ is derived from the Fourier transform of Eq.~\ref{eq:LLE} as:
\begin{equation}
t_p=\frac{i(\delta \omega_\text{eff}-\delta\omega_k) + (\kappa_i + \kappa_i' -\kappa_c )/2}{i(\delta\omega_\text{eff}-\delta\omega_k) + (\kappa_i + \kappa_i' +\kappa_c )/2},  \label{eq:pump}
\end{equation}
where $\delta\omega_k$ and $\kappa_i'$ are the additional frequency shift and conversion loss caused by the Kerr effect on the pump mode, respectively. Mathematically, they correspond to the real and imaginary parts of the Fourier series coefficients of the Kerr nonlinear term $F_p$ (i.e., $\delta\omega_k\equiv \text{Re}(F_p)$ and $\kappa_i'/2\equiv \text{Im}(F_p)$), given by:
\begin{equation}
F_p\equiv \gamma v_g  \frac{\mathcal{F}\left\{|E(t,\tau)|^2 E(t,\tau)\right\}_p}{\mathcal{F}\left\{E(t,\tau)\right\}_p}, \label{eq:Fp}
\end{equation}
where $\mathcal{F}$ denotes the Fourier transform and the subscript $p$ denotes the pump index. Physically, the additional loss $\kappa_i'$ stems from frequency conversion from the pump to other comb lines. Therefore, from conservation of energy we deduce that
\begin{equation}
\kappa_i'=(X-1)\kappa, \label{eq:ki}
\end{equation}
where $\kappa=\kappa_i + \kappa_c$ is the total loss rate and $X$ is the ratio of the intracavity power of the whole comb spectrum to that of the pump comb line (i.e., $X\equiv \sum_m |E_m|^2 /|E_p|^2$ with $E_m\equiv \mathcal{F}\left\{E(t,\tau)\right\}_m$).

\begin{figure*}[htbp]
	\centering
	\includegraphics[width=0.9\linewidth]{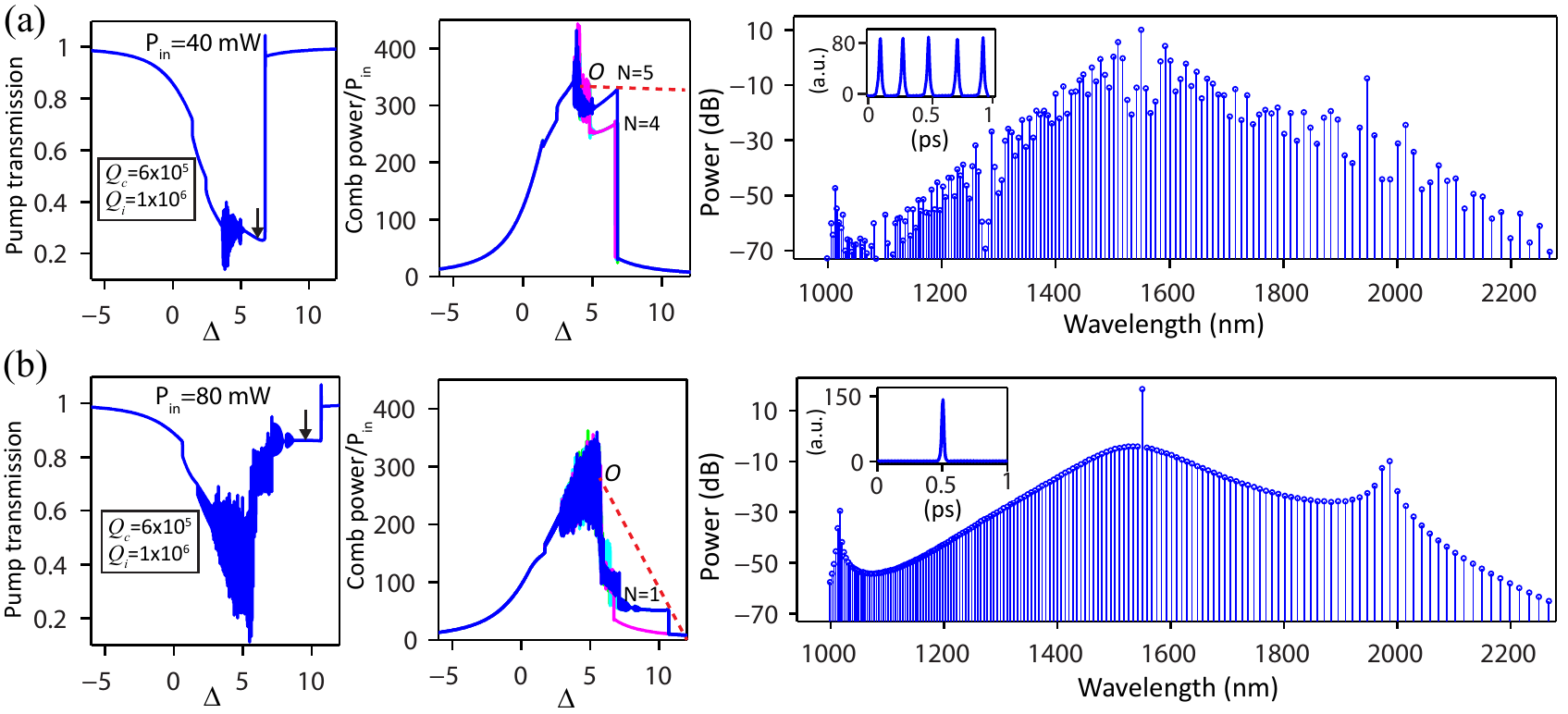}
	\caption{LLE simulation for a Si$_3$N$_4$ microring with thickness of 600 nm and ring width of 1760 nm for (a) pump power of 40 mW and for (b) pump power of 80 mW. In (a) and (b), from left to right we plot the pump transmission, the superimposed comb power for repeated simulations runs, and the comb spectrum when the pump detuning is set at the position marked by the vertical arrow in the pump transmission (the inset shows the corresponding temporal response). The absolute value of the slope of the red dashed lines in the central figures in (a) and (b) indicate the minimum thermal parameter $K_{\text{eff}}$ needed to stably access the $N=5$ soliton state in (a) and the $N=1$ soliton state in (b). A power of 0 dB is referenced to 1 mW (i.e., dBm).}
	\label{fig:sim_H600nm}
\end{figure*}

Figure~\ref{fig:thermal}(a) shows one example of numerical simulations based on the LLE model, with simulation parameters given in the Supplementary Material. For the moment we assume there is no thermal effect so the effective cavity detuning is solely determined by the laser frequency variation. By continuously varying $\delta\omega_\text{eff}$ in an increasing direction corresponding to decreasing the pump frequency, as done in experiments), we reproduce the three main stages of the comb generation: primary comb, MI comb and soliton. For convenience, the $x$ axis is normalized by the cavity half-linewidth ($\Delta\equiv 2\delta\omega_\text{eff}/\kappa$). In the lower panel of Fig.~\ref{fig:thermal}(a), we plot two normalized parameters which are defined as $\delta \tilde \omega\equiv 2(\delta\omega_\text{eff}-\delta\omega_k)/\kappa$ (blue line) and $\tilde\kappa_i'\equiv \kappa_i/\kappa$ (red line). The transition from the MI comb to the soliton is correlated with the sign change of $\delta\tilde\omega$, consistent with the expectation that solitons exist on the red-detuned side of the Kerr-shifted resonance~\cite{Kippenberg_soliton_2014}. Moreover, in the soliton region $\delta\tilde\omega$ is almost constant, suggesting that the Kerr phase shift can effectively compensate the cavity detuning to maintain the soliton form. Finally, the relationship specified by Eq.~\ref{eq:ki} is verified by comparing the numerical values of $X-1$ (black circles) with $\tilde \kappa_i'$ (red line) in the soliton region. Note that the value of $X$ is determined by the soliton state (which in turn depends on the pump power and dispersion) and typically increases as the soliton order reduces. For $X\gg1$, Eq.~\ref{eq:ki} indicates that the additional conversion loss dominates the coupling loss, and therefore, once the soliton is generated the pump mode is effectively in the under-coupled regime regardless of the initial coupling condition in the linear case.

 \begin{figure*}[t]
	\centering
	\includegraphics[width=0.9\linewidth]{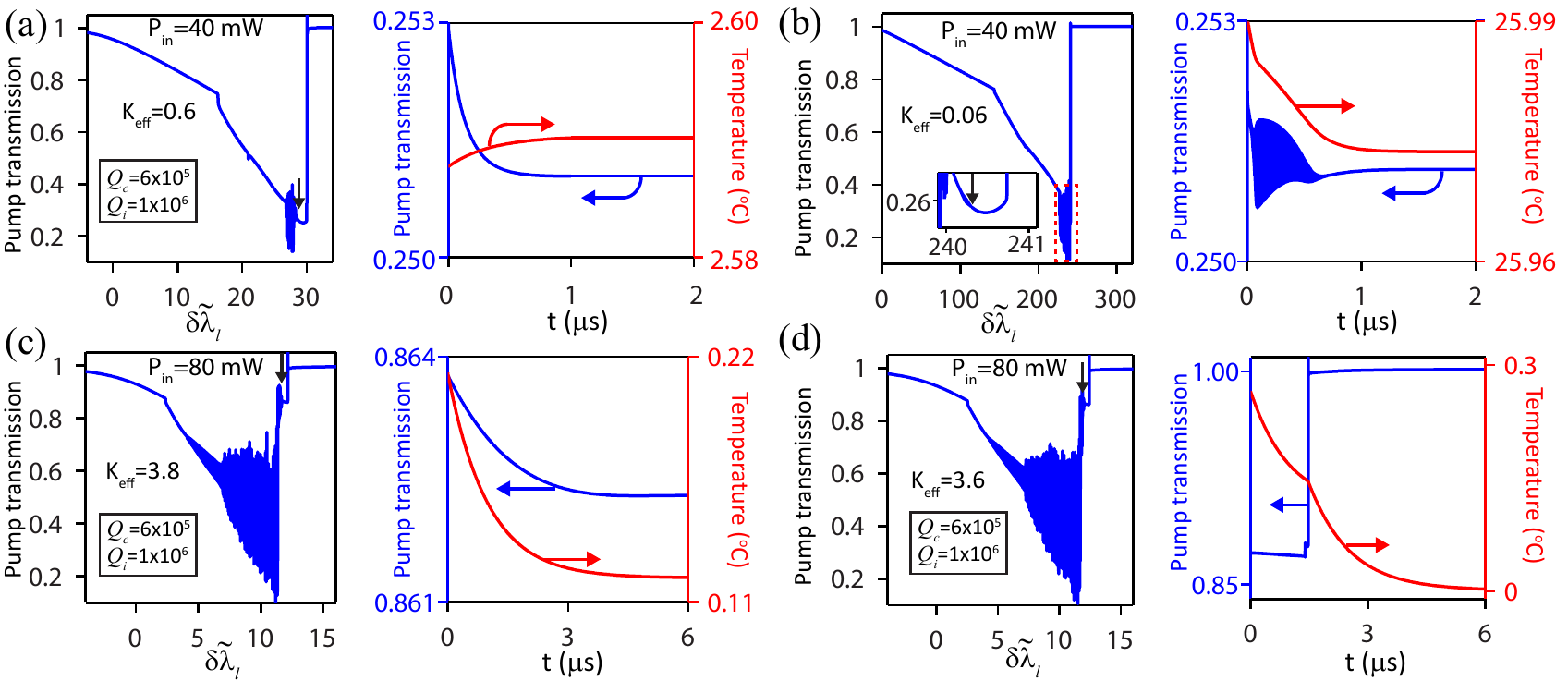}
	\caption{Full LLE simulation with the thermal effect included for the same Si$_3$N$_4$ microring studied in Fig.~\ref{fig:sim_H600nm}. The pump power and the effective thermal conductance $K_\text{eff}$ are varied for four cases: (a) pump power of 40 mW and $K_\text{eff}=0.6$; (b) pump power of 40 mW and $K_\text{eff}=0.06$; (c) pump power of 80 mW and $K_\text{eff}=3.8$, and (d) pump power of 80 mW and $K_\text{eff}=3.6$. For each case, in the left figure we show the pump transmission as a function of the laser wavelength detuning, which is gradually increased from the blue side of the pump resonance and its variation speed is slow compared to the thermal dynamics in the cavity. In the right figure we plot the temporal evolution of the pump transmission and the cavity temperature relative to the ambient environment by fixing the laser wavelength detuning at the position marked by the vertical arrow in the pump transmission.}
	\label{fig:fullLLE}
\end{figure*}

Next, we move on to discuss the impact of the thermal effect to the comb dynamics. First, we rewrite the effective cavity detuning as $\delta\omega_\text{eff}=(\omega_p-\omega_o)-(\omega_l-\omega_o)=\delta\omega_p -\delta\omega_l$, where $\omega_o$ is the cold-cavity resonance frequency, $\delta\omega_p$ is the resonance frequency shift due to the thermo-optic effect ($\delta\omega_p\equiv\omega_p-\omega_o$), and $\delta\omega_l$ is the laser frequency detuning relative to $\omega_o$ ($\delta\omega_l\equiv\omega_l-\omega_o$). Assuming a linear absorption of the incracavity power and the steady-state case, we have
\begin{equation}
\Delta=\delta\tilde\lambda_l -\frac{1}{K_\text{eff}}\sum_m |E_m|^2, \label{eq:thermal}  \\
\end{equation}
where $\delta\tilde\lambda_l$ is the normalized laser wavelength detuning ($\delta\tilde\lambda_l\equiv-2\delta\omega_l/\kappa$), and $K_\text{eff}$ is a parameter characterizing the resonance frequency shift caused by the thermal effect:
\begin{equation}
K_\text{eff}^{-1}\equiv 2\frac{\kappa_a}{\kappa}\frac{dn}{dT}\frac{\omega_0 t_R}{n_gK_c}, \label{eq:Keff}
\end{equation}
where $\kappa_a$ is the linear absorption rate, $\frac{dn}{dT}$ is the thermo-optic coefficient, $n_g$ is the group index, and $K_c$ is the thermal conductance of the microring resonator (see Supplementary Material).

We now incorporate the thermal effect into the problem following a two-step process: first we find the comb power ($\sum_m|E_m|^2$) as a function of the effective detuning ($\Delta$), and then we combine the obtained results with the thermal equation (Eq.~\ref{eq:thermal}) to find the solution for a given laser detuning $\delta\tilde{\lambda_l}$. This approach is illustrated in Fig.~\ref{fig:thermal}(b) using a graphical method. We start with a simple example which does not have the Kerr effect (inset of Fig.~\ref{fig:thermal}(b)). In this case, the cavity power is a Lorentzian function of the effective cavity detuning. On the other hand, Eq.~\ref{eq:thermal} specifies the cavity power as a linear function of the cavity detuning, with a slope of $-K_\text{eff}$ and $x$-intercept of $\delta\tilde{\lambda_l}$. Their intersecting points correspond to the steady-sate solutions. As we increase the laser wavelength, the number of solutions increases from one (line I) to multiple (between lines II and III) and reduces back to one again (beyond line III). The multiple-solution region is bounded by lines II and III, and at the intersecting points $\mathbf{O}$ and $\mathbf{N}$ (see inset of Fig.~\ref{fig:thermal}(b)), the tangent to the cavity power curve has the same slope of $-K_\text{eff}$ as specified by the thermal equation (Eq.~\ref{eq:thermal}). It is easy to deduce that the solution located between the points $\mathbf{O}$ and $\mathbf{N}$ (green line in the inset of Fig.~\ref{fig:thermal}(b)), where the slope of the tangent to the cavity power is smaller than $-K_\text{eff}$, is thermally unstable, in the sense that the temperature of the mode can be easily perturbed in this region to evolve to the thermally stable branch. In particular, if we further increase the laser wavelength at line III, the state of cavity will abruptly switch from the point $\mathbf{O}$ to the point $\mathbf{P}$, leading to a sharp increase in the pump transmission and resulting in the so-called thermal triangle as observed in the experiment (Fig.~\ref{fig:exp_H600nm}). We note that this analysis of the thermo-optic dispersion of a cavity is fully consistent with the literature \cite{Vahala_thermal}.

Having verified the graphical analysis in the case of no Kerr effect, we now apply the the same method to comb generation, with the following considerations. First, the comb power as a function of the effective cavity detuning is computed from the LLE simulation. Second, as can be seen from Fig.~\ref{fig:thermal}, the comb power in the unstable MI comb (chaotic) regime has strong oscillations. For the thermal analysis, since the thermal effect is much slower than the cavity lifetime, we only need to consider the averaged comb power. Finally, the transition from the unstable MI comb to the soliton is not a deterministic process, and soliton states with different orders can be accessed with a certain probability \cite{Kippenberg_backward_tuning}. For instance, Fig.~\ref{fig:thermal}(b) shows the possible soliton states for the microresonator operated under the same conditions as shown in Fig.~\ref{fig:thermal}(a). It is easy to see that for each soliton state there is a minimum $K_\text{eff}$ to make it thermally stable (i.e., otherwise there is no intersecting point), and the required $K_\text{eff}$ increases as the soliton number reduces. Moreover, this method provides a straightforward way to determine whether we can access certain soliton states in the experiment. Using Fig.~\ref{fig:thermal}(b) as an example, the maximum laser wavelength detuning for accessing $N=6$ (corresponding comb power shown by the blue line) is marked as $(\delta\tilde{\lambda}_l)_\text{max}$. If in the experiment the measured nonlinear triangle (as shown in Fig.~\ref{fig:exp_H600nm} in the pump transmission) is less than this value, we expect the $N=6$ and higher-order soliton states are thermally stable and can be accessed with the slow pump frequency tuning method.

\section{Thermal stability: Numerical simulations}
\label{sec:thermal_stability_sims}

We now use the two-step process outlined in the previous section to understand the available soliton states for the microresonator studied in our experiments, and whether they can be accessed with slow pump frequency tuning. As we show below, this approach provides a qualitative understanding of the experimental data. In addition, we compare our approach with a full LLE simulation that directly incorporates thermal effects, and we find that quantitative agreement can be achieved.

\begin{figure*}[t]
	\centering
	\includegraphics[width=0.9\linewidth]{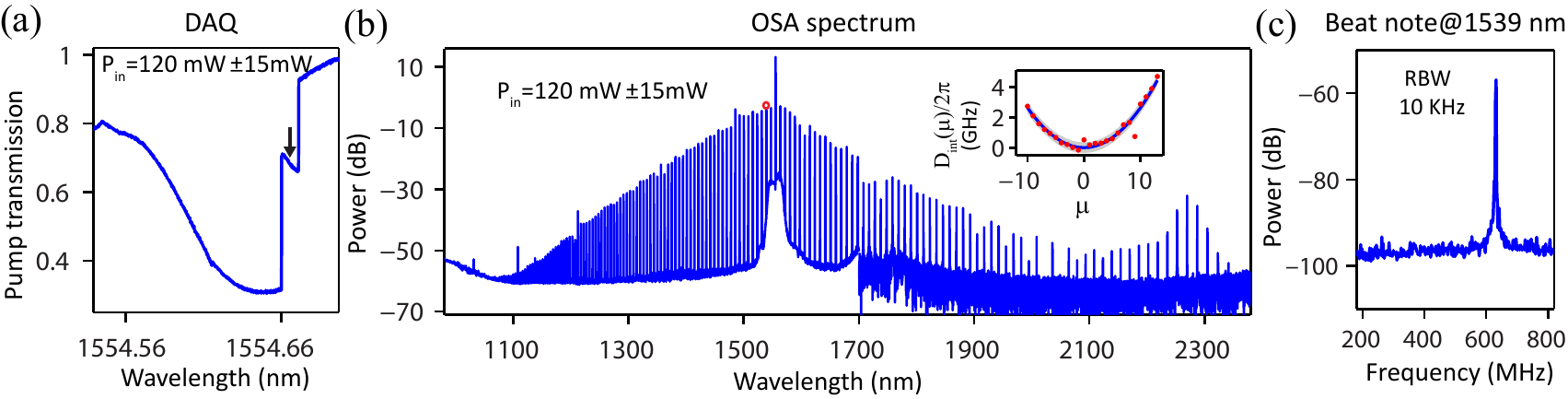}
	\caption{Experimental results for a Si$_3$N$_4$ microring with width of 1750 nm $\pm$ 10 nm and height of 623 nm $\pm$ 2 nm. (a) Pump transmission for the input power of 120 mW $\pm$ 15 mW, measured by the DAQ at a pump tuning speed of $-100$ GHz/s. (b) Comb spectrum corresponding to the soliton step with the pump detuning marked by the vertical arrow in (a). The inset shows the measured resonance dispersion (red dots) versus a quadratic fit (blue solid line), from which D$_2/2\pi$ = 53 MHz $\pm$ 2 MHz is extracted. (c) Beat-note measurement for the comb line near 1539 nm (marked by red circle in the spectrum in (b)). In (b) and (c), a power of 0 dB is referenced to 1~mW. The spectrum in (b) includes output fiber coupling loss ($\approx$6~dB at the pump wavelength).}
	\label{fig:exp_H623nm}
\end{figure*}

Figure~\ref{fig:sim_H600nm} shows the simulation results based on the LLE model for the microring studied experimentally in Fig.~\ref{fig:exp_H600nm}, with thickness of 600 nm and ring width of 1760 nm. The coupling/intrinsic $Q$s are estimated from the linear transmission measurement to be approximately $6\times 10^5/1\times10^6$, respectively. At the pump power of 40 mW (Fig.~\ref{fig:sim_H600nm}(a)), we observe a soliton step near the bottom of the pump transmission, which is similar to the experimental data shown in Fig.~\ref{fig:exp_H600nm}(a). By repeating the same simulation, we find the comb can end up in either the $N=5$ or $N=4$ soliton state. The similarity between the simulated comb spectrum for the $N=5$ soliton state (Fig.~\ref{fig:sim_H600nm}(a)) and the measured data (Fig.~\ref{fig:exp_H600nm}(b)) suggests that the multi-soliton state generated in the experiment is likely to be the $N=5$ soliton state. Moreover, using the graphical method illustrated in Fig.~\ref{fig:thermal}(b), we infer that the required $K_\text{eff}$ to make the $N=5$ soliton state thermally stable can be very small, since the slope of the intersecting line to the $N=5$ soliton state (red dashed line in the comb power plot in Fig.~\ref{fig:sim_H600nm}(a)) is close to zero. This is consistent with the experimental result that we are able to stably access such a multi-soliton state using slow pump frequency tuning.

Next, we compare the simulation results for the pump power of 80 mW to the experimental data shown in Fig.~\ref{fig:exp_H600nm}(c). The LLE simulation provided in Fig.~\ref{fig:sim_H600nm}(b) confirms that the high-order soliton states near the bottom of the pump transmission become chaotic with the increased pump power, and there is only one low-noise soliton step that is well above the transmission minimum. The simulated comb spectrum and temporal response shown in Fig.~\ref{fig:sim_H600nm}(b) reveal that this is a single soliton state. However, to stably access this soliton state the maximum nonlinear wavelength shift (contributed by the Kerr and thermo-optic effects) should be less than approximately 12 $\Delta$ (corresponding to a wavelength shift of 25 pm), which is considerably smaller than the experimentally observed thermal triangle with a wavelength span around 120 pm (see Fig.~\ref{fig:exp_H600nm}(c)). As a result, we conclude that this soliton state is not thermally stable due to the relatively strong absorption in our device, which is consistent with the experimental observation shown in Fig.~\ref{fig:exp_H600nm}(c).

Thus, our graphical method has achieved qualitative agreements with the experimental data in terms of the thermal stability of the available soliton state. We can further validate our approach by comparing its predictions with a full numerical simulation that has incorporated the thermal effect into the LLE model \cite{Kippenberg_soliton_2014, Kippenberg_backward_tuning}. For this purpose, the thermal decay rate of the Si$_3$N$_4$ microresonator ($\gamma_T\approx 2.9\times 10^5/s$) is obtained by implementing the heat transfer equation using a finite element method (details in the Supplementary Material). In addition, the laser detuning has been varied at a slow enough rate (relative to the thermal decay rate) to be consistent with the experiment (the scanning time is at least $100\gamma_T^{-1}$). Figure~\ref{fig:fullLLE}(a) shows such a simulation example, where $K_\text{eff}=0.6$ is chosen to match the experimentally measured thermal triangle at the pump power of 40 mW (see Fig.~\ref{fig:exp_H600nm}(a)). As can be seen, a soliton step has been reproduced near the bottom of the pump transmission as we gradually increase the laser wavelength from the blue side of the cavity resonance. Moreover, if we fix the laser detuning after it reaches the soliton step  (position marked by the vertical arrow in Fig.~\ref{fig:fullLLE}(a)) and let the comb state evolve by itself, we find that the temperature of the cavity and the pump transmission are both stabilized, thus demonstrating that this multi-soliton state is indeed thermally stable. Next, we reduce $K_\text{eff}$ by one order of magnitude, and find that the same soliton state can be stably accessed with slow frequency tuning of the pump laser (Fig.~\ref{fig:fullLLE}(b)). This agrees with the result from the graphical method in Fig.~\ref{fig:sim_H600nm}(a), which predicts that for $N=5$ soliton state the required $K_\text{eff}$ can be close to zero. Of course, the reduced value of $K_{\text{eff}}$ has some effect on the system, in elevating the steady-state cavity temperature much more above the ambient environment than in Fig.~\ref{fig:fullLLE}(a). In Figs.~5(c) and (d), we carry out similar full LLE simulations for the pump power of 80 mW, where we find $K_\text{eff}$ has to be larger than 3.6 to make the single soliton state thermally stable. It is straightforward to verify that a similar threshold for $K_\text{eff}$ can be obtained from the graphical method in Fig.~\ref{fig:sim_H600nm}(b). Therefore, we conclude that the two-step analysis developed in this work provides a convenient and accurate approach to analyze the thermal stability of soliton states.

\section{Accessing the single soliton state}
\label{sec:single_solitons}

So far we have shown, both experimentally and theoretically, that we can access a high-order soliton state with properly chosen pump powers. In addition, we have shown that it could be quite challenging to access low-order soliton states with slow pump frequency tuning, as the comparatively larger drop in intracavity comb power when crossing from the MI regime to the soliton regime places a more stringent requirement on the effective thermal property of the microresonator. For instance, the simulation performed in Fig.~\ref{fig:fullLLE} indicates that we need to increase $K_\text{eff}$ by a factor of six to stably access the single soliton state in our microresonator. From the definition of $K_\text{eff}$ (Eq.~\ref{eq:Keff}), this can be done by reducing the absorption rate or increasing the thermal conductance of the resonator, for example. However, in the experiment we find that it is possible to access the single-soliton state in some devices without significant improvements to any of these properties. We provide one such example in Fig.~\ref{fig:exp_H623nm}, which is measured from a microring with thickness of 623 nm $\pm$ 2 nm and ring width of 1750 nm $\pm$ 10 nm. As shown in Fig.~\ref{fig:exp_H623nm}(a), a thermally stable soliton step is observed for a pump power of 120 mW $\pm$ 15 mW, and the corresponding soliton state can be accessed through slow pump frequency tuning. Interestingly, this thermally stable soliton step only appears in the DAQ transmission if the input pump has a mixed polarization configuration, and disappears if the polarization is adjusted to be pure TE (transverse electric). In fact, we have observed similar behavior (single soliton generation being accessed for slow frequency tuning with a mixed polarization pump) for multiple devices across multiple chips. The comb spectrum provided in Fig.~\ref{fig:exp_H623nm}(b) implies that this is a single-soliton state, which spans more than one octave and has a long dispersive wave around $2.27 \ \mu$m. Its coherence is verified by the beat-note measurement shown in Fig.~\ref{fig:exp_H623nm}(c). Finally, the measurement of the resonance frequencies in the $1.55\ \mu$m band (see the inset of Fig.~\ref{fig:exp_H623nm}(b)) gives $D_2/2\pi=53$ MHz $\pm$ 2 MHz. The stronger dispersion than that of the microring studied in Fig.~\ref{fig:exp_H600nm} is attributed to the thicker Si$_3$N$_4$ layer and the slightly smaller ring width. It is also consistent with the fact that we only see one long dispersive wave in the comb spectrum (see Fig.~\ref{fig:setup}(a)).

\begin{figure*}[t]
	\centering
	\includegraphics[width=0.9\linewidth]{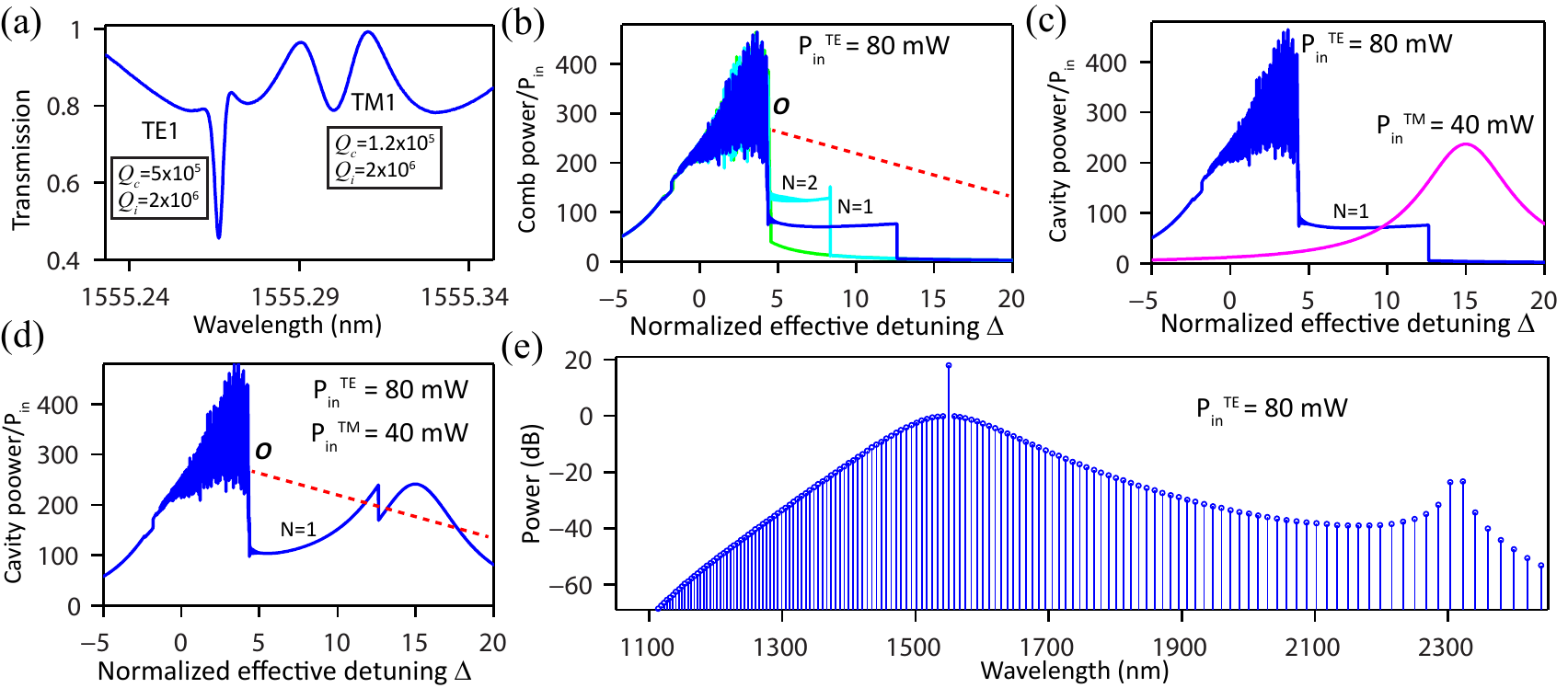}
	\caption{Thermal stability analysis for the microring studied in Fig.~\ref{fig:exp_H623nm}. (a) Experimental linear transmission measurement showing the adjacent TM mode on the red-detuned side of the TE mode. Their intrinsic/coupling quality factors are also provided. (b) LLE simulation for a fixed pump power of 80 mW and a blue frequency shift equal to half of the cavity linewidth ($\kappa/2$) at the pump mode. Different colors represent possible outcomes from repeated runs of the same LLE simulation. The red dashed line has a slope ($-K_\text{eff}$) inferred from the experimentally measured thermal triangle in Fig.~\ref{fig:exp_H623nm}(a), and has no intersecting point with the single soliton step. (c) Simulated cavity power for the TE (blue) and TM (magenta) modes with an input power of 80 mW and 40 mW, respectively. The effective detuning is referenced to the cold cavity resonance of the TE mode and normalized by its half-linewidth. The spectral position of the TM mode is determined from the linear transmission shown in (a). (d) Calculated total cavity power from (c). The red dashed line is the same as in (b), and now has an intersecting point with the single soliton step. (e) Simulated comb spectrum corresponding to the single soliton state for an input power of 80 mW. Here, a power of 0 dB is referenced to 1mW.}
	\label{fig:sim_H623nm}
\end{figure*}

Before trying to explain the experimental results, we first notice that there is an avoided mode crossing at the pump resonance, which leads to a blue frequency shift in the pump resonance as evident in the dispersion measurement in Fig.~\ref{fig:exp_H623nm}(b) (where the pump resonance frequency is blue shifted nearly half of the cavity linewidth away from the origin of a quadratic dispersion fit). We identify from the linear transmission measurement (Fig.~\ref{fig:sim_H623nm}(a)) that this frequency shift is caused by the mode interaction between the fundamental TE and TM (transverse magnetic) modes \cite{Gaeta_TETM_coupling} (note that the fundamental TM mode has normal dispersion in the 1550 nm band and is severely overcoupled). The LLE simulation performed in Fig.~\ref{fig:sim_H623nm}(b) suggests that the amount of the frequency shift of the pump resonance due to the TE-TM mode coupling is tolerable, and the soliton generation is not inhibited, as has been seen in other works in which mode crossings near the pumped mode are observed \cite{Kippenberg_mode_crossing}. However, using the same graphical method developed in Sect.~5, we find that the maximum nonlinear wavelength shift corresponding to a thermally stable single soliton state is about 35 pm. This is considerably less than the typical thermal triangle observed in the experiment, which is on the order of 100 pm for a pump power of 80 mW (see Fig.~\ref{fig:exp_H600nm}(c) and Fig.~\ref{fig:exp_H623nm}(a)). This explains the fact that soliton step is absent in the DAQ transmission when the input polarization is set to pure TE. In addition, we note that in our experiments, stable soliton generation is not observed when we pump on nearby TE modes (i.e., those which do not exhibit TE-TM mode coupling).

Thus, it seems from experiments and initial simulations that both the mixed polarization input and the presence of a nearby TM mode are needed for stable single soliton generation in these devices. We now try to understand how these elements help in accessing stable single soliton states through slow pump frequency tuning. It is important to realize that though the two modes shown in Fig.~\ref{fig:sim_H623nm}(a) are considered to be independent in the linear case, they can interact with each other at high pump powers through the thermal effect. For example, the absorption of the intracavity power from the TE mode leads to a rise of the cavity temperature, which in turn shifts the resonance frequency of the TM mode. The same is true with the TM mode. In our case, the presence of a second mode (TM) on the red-detuned side of the pump resonance (TE) could help stabilize the thermal dynamics as the frequency comb evolves to the soliton state. While previous studies have shown that in some cases the avoided mode crossing could help the comb formation in terms of the Kerr dynamics \cite{Maleki_mode_crossing, Weiner_mode_crossing}, here the benefit from the adjacent mode is purely from the consideration of the thermal dynamics.  To see this, we use Fig.~\ref{fig:sim_H623nm}(c) as one example, where we have plotted the intracavity power for both the TE and TM modes for input powers of 80 mW and 40 mW, respectively. Because of the relatively small input power as well as the low Q factor, the intracavity power of the TM mode can be approximated as a Lorentzian function of the effective cavity detuning (i.e., we neglect the Kerr effect for the TM mode). The sum of the cavity power from the two modes (assuming that the TE and TM modes have the same absorption rate), which determines the cavity temperature, is plotted in Fig.~\ref{fig:sim_H623nm}(d). Using the same graphical method as in the previous section, we find that the single soliton state becomes thermally stable for $K_\text{eff}$ that matches the experiment. Notably, this single soliton state is not thermally stable if the TM mode is absent (Fig.~\ref{fig:sim_H623nm}(b)). Together, this demonstrates that an adjacent mode on the red-detuned side of the pump resonance can mitigate the thermal requirements for accessing low-order soliton states. While these modes can in principle be of the same polarization, orthogonal polarization enables a level of in-situ control of the thermal dynamics, as adjustment of the input polarization state varies the relative power coupled into each mode. Finally, the simulated comb spectrum corresponding to the single soliton state for an input power of 80 mW for the TE mode is provided in Fig.~\ref{fig:sim_H623nm}(e), and a reasonably good agreement with the experimental data shown in Fig.~\ref{fig:exp_H623nm}(c) is observed.

\section{Conclusion}
The successful demonstration of phase-coherent, octave-spanning soliton microcomb states in a chip-integrated Si$_3$N$_4$ microresonator represents an important step towards on-chip self-referencing \cite{Kippenberg_self_ref1,Kippenberg_self_ref2}. In particular, the use of a relatively small pump power (approximately 100 mW or less) and the possibility of generating double dispersive-wave emission are two promising features for many practical applications. Our soliton microcomb states can be accessed with slow frequency tuning of the pump laser, thus relinquishing the need for more complicated control of pump power and/or frequency. More generally, we have presented a systematic treatment analyzing the thermal stability of soliton states, from which we conclude that high-order soliton states are relatively easy to generate by carefully choosing the pump power, so that they are thermally stable and not in the chaotic regime. In contrast, accessing lower-order/single soliton states with slow pump frequency tuning could be challenging and may require improved resonator properties.  This could include reducing the optical absorption rate (e.g., through improved annealing and surface treatments to reduce bulk and surface absorption) or engineering the device geometry to increase the resonator thermal conductance. Alternatively, we also show that it is possible to overcome the thermal challenges by taking advantage of mode couplings which do not sufficiently perturb the mode spectrum to inhibit soliton formation, but which modify the thermal landscape of the system.  This method shows the possibility that with proper thermal engineering of the microresonator, accessing the phase-coherent single soliton microcomb states could become a straightforward process in the future.

\section{Supplementary Material: Simulations on thermal parameters}
The thermal dynamics in a microresonator are described by the following equation~\cite{ilchenko1992thermal,Vahala_thermal}:
\begin{equation}
\frac{d T_\text{eff}}{dt}= -\gamma_T\left(T_\text{eff} - \frac{P_\text{abs}}{K_c}\right), \label{eq:dT}
\end{equation}
where $\gamma_T$ is the thermal decay rate, $K_c$ is the thermal conductance of the microresonator, and $P_\text{abs}$ denotes the absorbed optical power. $T_\text{eff}$ is an effective temperature computed by averaging the temperature of the cavity, $T(\textbf{r},t)$, over the optical mode volume as
\begin{equation}
T_\text{eff}=\frac{\int T(\textbf{r},t) n^2(\textbf{r}) |E(\textbf{r})|^2\ d^3\textbf{r}}{\int n^2(\textbf{r}) |E(\textbf r)|^2 \ d^3 \textbf{r}},  \label{eq:Teff}
\end{equation}
where $n(\textbf{r})$ denotes the refractive index and $E(\textbf{r})$ is the electric field of the cavity mode.

\begin{figure*}[htbp]
	\centering
	\includegraphics[width=0.9\linewidth]{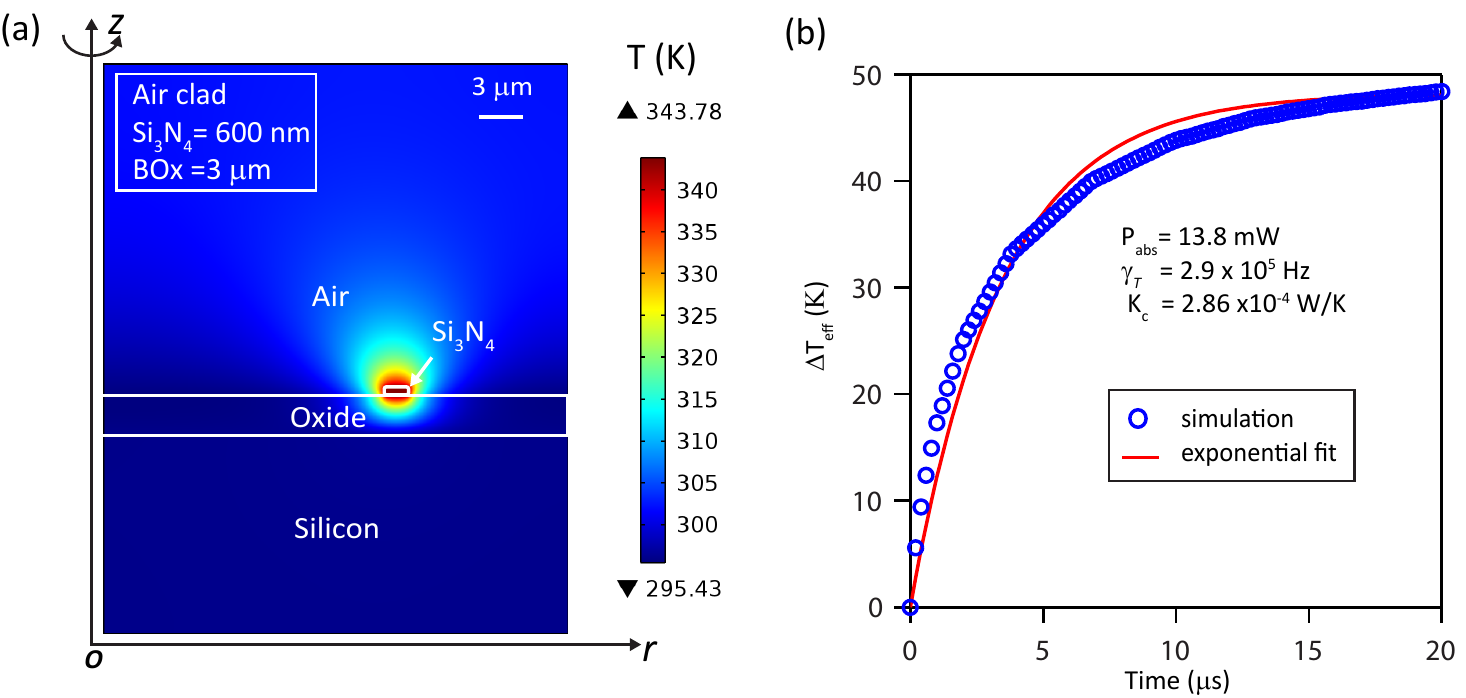}
	\caption{Numerical simulation of the thermal properties for an air-clad Si$_3$N$_4$ microresonator. (a) Steady-state temperature distribution of a 23 $\mu$m radius Si$_3$N$_4$ microring resonator with $13.8$ mW absorption power. The Si$_3$N$_4$ thickness is 600 nm and the buried oxide (BOx) layer thickness is 3 $\mu$m. (b) Simulated effective temperature $T_\text{eff}$ relative to the ambient temperature (blue circles) after turning on the heat source at $t=0$. The red solid line is the exponential fit to extract the thermal decay rate and the thermal conductance of the resonator.}
	\label{fig:thermal_SM}
\end{figure*}

To obtain the numerical values of the the thermal decay rate and the thermal conductance, we implement the heat transfer equation based on the finite element method (FEM) for the resonator structure (See Fig.~\ref{fig:thermal_SM}(a)). The heat source is assumed to have the same intensity distribution as the resonant mode (i.e., we assume a linear absorption of optical power). By fitting the simulation results (Fig.~\ref{fig:thermal_SM}(b)) with Eq.~\ref{eq:dT}, we obtain $\gamma_T=2.9\times 10^5$ Hz and $K_c=2.86\times 10^{-4}$ W/K for a 23 $\mu$m radius Si$_3$N$_4$ microring resonator.

\section{Supplementary Material: Parameters used in LLE simulation}
In this section, we list the major parameters used in the LLE simulations shown in the main text, including Figs.~4-6 and 8. In Fig.~4, we have adopted similar parameters as used in Ref.~\cite{Kippenberg_backward_tuning}, where the dispersion of the microring is dominated by the second order dispersion ($\beta_2$). In all the other LLE simulations, the dispersion is obtained from a fully vectorial microresonator eigenfrequency mode solver based on FEM, which includes the bending dispersion present in THz mode spacing resonators (not significant for the 100~GHz mode spacing in Fig.~4), and higher-order dispersion terms are retained. The full LLE simulation shown in Fig.~6 is performed by combining Eq.~\ref{eq:dT} with the standard LLE model, using the thermal parameters obtained in Section 1. We assume a value of the Kerr nonlinear refractive index $n_{2}\approx2.5\times 10^{-19}$~m$^2$W$^{-1}$ for Si$_3$N$_4$~\cite{ikeda2008thermal,Lipson_OPO_nphoton2010}, and the effective nonlinearity $\gamma$ is determined from this value and the FEM-determined effective modal area~\cite{lin2007nonlinear,Kartik_FWMBS}.

\begin{table*}[ht]
\centering
\captionsetup{justification=centering}
\caption{Parameters used in the LLE simulations in the main text}
\begin{tabular}{|c|c | c|c|c|}
\hline
Figure No.& $Q_c/Q_i$ & Dispersion & Radius($\mu$m) &  Power (mW)   \\ [0.5ex]
\hline
4 & $1\times10^6/1\times 10^6$& $\beta_2=-1.6\times10^{-25}$ ps$^2$/(nm $\cdot$ km) (From Ref.~\cite{Kippenberg_backward_tuning}) & 230 & 750\\
5 \& 6 & $6\times10^5/1\times 10^6$& FEM simulation & 23 & 40, 80 (see the legend) \\
8 & $5\times10^5/2\times10^6$ & FEM simulation & 23 & 80 \\
\hline
\end{tabular}
\label{List_of_parameters}
\end{table*}

\section*{Funding Information}
Q.~Li acknowledges support under the Cooperative Research Agreement between the University of Maryland and NIST-CNST (award no. 70NANB10H193). This project is partially funded by the DARPA DODOS and NIST-on-a-Chip programs, as well as the Air Force Office of Scientific Research under award number FA9550-16-1-0016.

\section*{Acknowledgments}
The authors would like to acknowledge helpful discussions with Dr.~Karen Grutter, Dr.~Marcelo Davan\c co, Dr. Mohammad Soltani, and Dr. Jared Strait.  The device layout was done with the 'Nanolithography Toolbox', a free software package developed by the NIST Center for Nanoscale Science and Technology~\cite{nanolithography_toolbox_2016}.
%

\end{document}